\documentclass[preprint,showpacs,preprintnumbers,amsmath,amssymb]{revtex4}

\usepackage{graphicx}
\usepackage{dcolumn}
\usepackage{bm}

\usepackage{amssymb}
\usepackage{latexsym}
\usepackage{amsthm}
\usepackage{graphicx}
\theoremstyle{plain}

\newtheorem{theorem}{Theorem}
\newtheorem{lemma}{Lemma}
\newtheorem{corollary}{Corollary}
\newtheorem{proposition}{Proposition}

\begin{document}

\title{Limit theorems for quantum walks driven by many coins}

\author{Etsuo Segawa}
\email{segawa820@npde.osu.sci.ynu.ac.jp}
\affiliation{%
Department of Applied Mathematics, 
Yokohama National University, 
79-5 Tokiwadai, Yokohama, 240-8501, Japan\\}

\author{Norio Konno}
\email{norio@mathlab.sci.ynu.ac.jp}
\affiliation{%
Department of Applied Mathematics, 
Yokohama National University, 
79-5 Tokiwadai, Yokohama, 240-8501, Japan\\}




\begin{abstract}
We obtain some rigorous results on limit theorems for quantum walks driven by many coins 
introduced by Brun \textit{et al.} in the long time limit. 
The results imply that whether the behavior of a particle is quantum or classical 
depends on the three factors: 
the initial qubit, the number of coins $M$, 
$d= [t/M]$, where $t$ is time step. 
Our main theorem shows that we can see a transition from classical behavior to quantum one 
for a class of three factors.
\end{abstract}

\maketitle


\section{Introduction}	
\quad Discrete-time quantum walks are considered as quantum counterparts of 
discrete-time classical random walks \cite{Konno book,KonnoA2,Meyer}. 
It is often showed that a useful quantum search algorithm can be designed based on the quantum walk \cite{Ambainis2}. 
Moreover, the result on a quantum walk on $\mathbb{Z}$ with an absorbing wall \cite{Konno ab}
has been applied to solve the transport problems in 
solid-state physics of strongly correlated electron systems in \cite{Oka}.\\
\quad 
The important factor describing a difference between the quantum walk and the classical random walk is 
the order of variance in the long time limit. 
Let $X_t$ and $\widetilde{X}_t$ describe the position on $\mathbb{Z}$ of finding a particle for 
a symmetric Hadamard walk and 
a symmetric classical random walk, respectively. 
To get the property appearing to the quantum walk, 
we will consider the limit theorem of weak convergence. 
The central limit theorem implies $\widetilde{X}_t/\sqrt{t}\Rightarrow \mathrm{N}(0,1)$ as $t \to \infty$, 
where $\mathrm{N}(0,1)$ is Gaussian with mean $0$ and variance $1$ 
and ``$\Rightarrow$'' means weak convergence, while 
$X_t/t \Rightarrow \mathcal{Z}$ as $t \to \infty$ with the following density function 
(see  \cite{KonnoA0,KonnoA1,Grimmett} for more details): 
\begin{equation}\label{Konnodens} 
\rho(x)=\frac{I_{(-1/\sqrt{2},1/\sqrt{2})}(x)}{\pi (1-x^2)\sqrt{1-2x^2}}, 
\end{equation}
where $I_A(x)$ is the indicator function of a set $A\subset \mathbb{R}$. 
Thus a quantum particle can travel to quadratically farther position than a classical particle on $\mathbb{Z}$. 
This property also suggests the usefulness of quantum walks for spatial quantum searches. \\
\quad In this paper, we show a transition from the quantum walk to the classical random walk 
in the view point of the scaling 
order for the weak convergence with respect to time steps. We treat the quantum walk 
introduced by Brun \textit{et al.} \cite{Brun0,Brun}. 
If we apply the $M$ coins to the quantum walk, the quantum walk is called as an \textit{$M$-coin quantum walk ($M$-CQW)} 
here. 
The $M$-CQW is a quantum generalization of the random walk depending 
on the previous $M$-step memory \cite{KonnoA2,Flitney}. 
The quantum coin describing the one step dynamics can be obtained 
by replacing the nonzero entries of an adjacency matrix of the de-Bruijn digraph to some nonzero appropriate values. 
See \cite{Flitney,Severini2} for more details. 
Throughout this paper, the initial qubit is given by two cases, i.e., (A) the initial qubit 
described by $M$-th tensor product of $\varphi_0$ with $||\varphi_0||^2=1$ and (B) the initial qubit chosen 
from the basis of coin space randomly. 
Brun \textit{et al.} \cite{Brun} showed the first and second moments of $M$-CQW in the long time limit 
with the initial qubit (A). 
We call \textit{arcsine law distribution} the distribution corresponding to arcsine law whose 
distribution has the following density function:
\begin{equation}\label{arcsine} 
s(x)=\frac{I_{(-1,1)}(x)}{\pi\sqrt{1-x^2}}. 
\end{equation} 
We often use the arcsine law distribution to describe the quantum property, 
for example, the distribution appears as the limit density function for continuous-time quantum walks 
on $\mathbb{Z}$\cite{Konno conti1,Konno conti2} and on hypercube \cite{Konno conti3}. 

Our main result provides the limit distribution for large time steps given by convolution of 
$\mathrm{N}(0,1)$ and arcsine law distribution on an appropriate setting, and 
it also provides the limit distribution of 
product of two independent random variables $\mathcal{X}$ and $\mathcal{Z}$ 
on another appropriate setting, 
where $\mathcal{X}$ has the distribution $\mathrm{N}(0,1)$ and $\mathcal{Z}$ has the distribution with 
the density function $\rho(x)$. 
\\
\quad 
This paper is organized as follows. Section 2 is devoted to the definition of $M$-CQW. 
In Section 3, we introduce a 
useful lemma to obtain the characteristic function of the position of finding a particle, and we give 
a proposition in relation to the weak limit theorems as time step $t\to \infty$ with $t=0\;(\mathrm{mod}\;M)$ 
by using the lemma. 
Section 4 proposes three types to find the quantum-to-classical transition and gives the limit theorems. 
At first, we consider the case $M<t<2M$ for large $M$. Secondly, we treat the other initial condition, that is, 
a mixture of cases (A) and (B). Finally, we give the weak limit theorems for 
$M=\Theta(t^{\alpha})$, $d=\Theta(t^{1-\alpha})$ $(0<\alpha<1)$ as $t \to \infty$, where we say that 
$\Theta(f(t))=g(t)$ if and only if $0<\lim_{t\to\infty}|f(t)/g(t)|<\infty$. 
\section {Definition of $M$-CQW}
The discrete-time quantum walk on $\mathbb{Z}$ is a quantum generalization of the random walk 
with an additional coin state $\mathcal{H}_C$. 
The state space is described by direct production of $\mathcal{H}_P$ (position state) and $\mathcal{H}_C$ (coin state). 
Here, $\mathcal{H}_P$ is associated with standard basis $|x\rangle$, $x \in \mathbb{Z}$. 
$\mathcal{H}_C$ is generated by chiralities corresponding to the directions of the motion of the particle. 
The one step unitary transformation $U$ is described by $U=S\cdot C$ with two operations, 
``coin flip'' $C$ and ``shift'' $S$. 
The coin flip operates on the coin state by using a unitary coin $H$ and leaves the position state. 
The shift operator moves the particle to one unit following the chirality of the particle. 
Let $|1\rangle_C={}^T[1,0]$, $|\!\!-\!\!1\rangle_C={}^T[0,1]$, where $T$ means the transposed operator. 
In $M$-CQW, the standard basis of the coin state is given by 
$|\xi_{M-1}\rangle_C \otimes \cdots \otimes |\xi_0\rangle_C\equiv |\xi_{M-1}, \dots, \xi_0\rangle$ with $\xi\in\{-1,1\}$. 
We prepare $M$ quantum coins $H_0,\;H_1,\dots,H_{M-1}$, where $H_j$ is given by 
$H_j=I^{^\otimes M-j-1}\otimes H \otimes I^{\otimes j}$ and $H$ is the two-dimensional Hadamard matrix:
\[ H=\frac{1}{\sqrt{2}}\begin{bmatrix}1&1\\1&-1\end{bmatrix}. \] 
At time $t$, the coin $H_{\mathrm{mod}[t,M]}$ acts on the coin state, 
where $\mathrm{mod}[t,M]$ means remainder of $t/M$:
\begin{equation}
C|x,\xi_{M-1},\dots,\xi_j,\dots,\xi_0\rangle 
		= |x,\xi_{M-1},\dots,H\xi_j,\dots,\xi_0\rangle. \label{half-step}
\end{equation}
where $j=\mathrm{mod}[t,M]$. 
$S$ moves the particle to one unit following the $j$-th chirality: 
\begin{equation}
S|x,\xi_{M-1},\dots,\xi_j,\dots,\xi_0\rangle = |x+\xi_j, \xi_{M-1},\dots,\xi_j,\dots,\xi_0\rangle. \label{one-step}
\end{equation}
We can express $U$ as 
\begin{equation}\label{one-step}
U= \widehat{S}\otimes \left(I^{^\otimes M-j-1}\otimes P_1 \otimes I^{\otimes j}\right)+
	\widehat{S}^{-1}\otimes \left(I^{^\otimes M-j-1}\otimes P_{-1} \otimes I^{\otimes j}\right), 
\end{equation} 
with $\widehat{S}=\sum_{x\in \mathbb{Z}}|x+1\rangle\langle x|$ and $P_j=|j\rangle\langle j|H$ $(j=1,-1)$. 
Remark that if $M=1$, the quantum walk is equivalent to $2$-state Hadamard walk on $\mathbb{Z}$. 
Assume that the time step is described by $t=dM+q$ with $0 \leq q <M$. 
Let $\Phi_{t}^{[M]}(x)$ be the coin state of a particle at time $t$ and position $x$. 
Let $X_t^{[M]}$ describe the position of finding a particle at time $t$. 
The probability of finding a particle at time $t$ and position $x$ is defined by 
\[ P(X_t^{[M]}=x)=||\Phi_t^{[M]}(x)||^2. \]
The spatial Fourier transformation of $\Phi_{t}^{[M]}(x)$ is given by 
$\widehat{\Phi}_{t}^{[M]}(k)=\sum_{x\in \mathbb{Z}}\Phi_t^{[M]}(x)e^{ikx}$. 
Assume the initial qubit is $\bigotimes_{j=1}^{M} \varphi_j$ 
with $||\varphi_j||^2=1$. Then from Eq. (\ref{one-step}), we have 
\begin{equation} \label{deB}
\widehat{\Phi}_{t}^{[M]}(k)=\bigotimes_{j=q+1}^{M}\widehat{\Psi}_{d}^{(\varphi_j)}(k) \otimes 
						\bigotimes_{j=1}^{q}\widehat{\Psi}_{d+1}^{(\varphi_j)}(k), 
\end{equation}
where $\widehat{\Psi}_d^{(\varphi_j)}(k)=\widehat{H}^d(k)\varphi_j$ with 
$\widehat{H}(k)=(e^{ik}|1\rangle\langle 1|+e^{-ik}|-1\rangle\langle -1|)H$. \\
\quad Miyazaki \textit{et al.} \cite{Katori} 
described $M$-CQW by their quantum walk with a quantum coin 
described by the $(2j+1)$-dimensional unitary representation of the rotation operator with half-integer $j$. 
The $M$-th tensor-product of $\mathbb{C}^2$ space $\mathcal{W}_M$ is decomposed into 
$(2j+1)$-dimensional irreducible rotation group modules $\mathcal{V}_j$ with half-integer $j$;
$\mathcal{W}_M\cong \bigoplus_{j}d_j^{[M]}\mathcal{V}_j$   
with 
\[d_j^{[M]}= \binom{M}{(M-2j)/2}-\binom{M}{(M-2j)/2-1}.\] The multiplicity $d_j^{[M]}$ is obtained by 
using the highest weight decomposition (see p.66 in \cite{georgi}). 
Thus the limit density function of $M$-CQW is described by a linear combination of their density functions. 
However it seems to be complicated to compute the unitary basis transformation operator 
$K_M:\mathcal{W}_M \to \bigoplus_j d_j^{(M)}\mathcal{V}_j$, so in this paper, we treat $M$-CQW directly. 
\section{The characteristic function for $M$-CQW }
\noindent
We consider the following two cases for the initial qubit $\Phi_0$; \\

\noindent
\textbf{Case (A)}:  
$\Phi_0=\varphi_0^{\otimes M}$, with $\varphi_0={}^T[1/\sqrt{2},i/\sqrt{2}]$. \\ 
\\
\textbf{Case (B)}: 
$\Phi_0=\phi_{M-1}\otimes \cdots \otimes \phi_0$, where $\{\phi_j\}_{j=0}^{M-1}$ is an i.i.d. sequence of 
Bernoulli random variables with $\textrm{Pr}(\phi_j=1)=\textrm{Pr}(\phi_j=-1)=1/2$. \\

Remark that the initial qubit of Case (A) is in a pure state, while the initial qubit of Case (B) is 
in a mixed state \cite{N-C}. 
Brun \textit{et al.} \cite{Brun} computed the first and second moments of $M$-CQW 
with the initial qubit of Case (A) in the long time limit and 
showed that the variance grows in proportion to square time steps.  
By definition of the spatial Fourier transformation of $\Psi_d^{(M)}(x)$, we have 
\begin{equation} 
E\left(e^{i\xi X_{t}^{[M]}}\right)
	=\int_{0}^{2\pi} \langle \hat{\Phi}_t^{[M]}(k),\;\hat{\Phi}_t^{[M]}(k+\xi)\rangle \frac{dk}{2\pi}. 
\end{equation}
Then applying Eq. (\ref{deB}) to the above equation, 
we can provide the following lemma with respect to the characteristic function for $X_t^{[M]}$ 
to obtain some weak convergence theorems: 
\begin{lemma}\label{chara} 
Let $Q_d(k,\xi)=\langle \widehat{\Psi}_d^{(\varphi_0)}(k), \widehat{\Psi}_d^{(\varphi_0)}(k+\xi)\rangle$ and 
$C_d(k,\xi)=\mathrm{Tr}[\widehat{H}^{-d}(k)\cdot\widehat{H}^{d}(k+\xi)]/2$. 
Assume that $t=dM+q$ with $0\leq q<M$. 
Then the characteristic function of $X_t^{[M]}$ can be expressed as \\
\textbf{Case (A):}
\begin{equation}
E\left(e^{i\xi X_{t}^{[M]}}\right)=\int_{0}^{2\pi}Q_{d+1}(k,\xi)^q Q_{d}(k,\xi)^{M-q}\frac{dk}{2\pi}.
\end{equation}
\textbf{Case (B):}
\begin{equation}
E\left(e^{i\xi X_{t}^{[M]}}\right)=\int_{0}^{2\pi}C_{d+1}(k,\xi)^q C_{d}(k,\xi)^{M-q}\frac{dk}{2\pi}.
\end{equation}
\end{lemma}
By Eq. (\ref{one-step}), we have 
$\Phi_t^{[M]}(x)=\sum_{\eta_M+\cdots+\eta_1=x}P_{\eta_M}\otimes \cdots \otimes P_{\eta_1}\Phi_0$ for $t=M$. 
Note that $\langle P_i\eta, P_j\eta\rangle=\delta_{ij}/2$ for $\eta\in\{\phi_0, |-1\rangle, |1\rangle\}$. Therefore 
$X_t^{[M]}$ has a binomial distribution $B(M,1/2)$. 
\begin{proposition}\label{prop1}
For both Cases (A) and (B), $X_t/\sqrt{t} \Rightarrow \mathrm{N}(0,1)$ as $t \to \infty$ with $t\leq M$. 
\end{proposition}
In the rest of this section, we will show a weak limit theorem taking $t \to \infty$ with $q=0$, i.e., $t=dM$. 
Put 
\[ \rho(x)=\frac{I_{(-1/\sqrt{2},1/\sqrt{2})}(x)}{\pi(1-x^2)\sqrt{1-2x^2}}, \]
which is the density function for the symmetric Hadamard walk \cite{KonnoA0,KonnoA1,Grimmett} in the long time limit. 
\begin{proposition}\label{prop2}
Assume that $t=dM$. 
\begin{enumerate}
\item \label{th-Mfix}If we fix $M$ and take $d\to \infty$, then 
with the initial qubit of Case (A) (resp. (B)), 
$X_t^{[M]}/t\Rightarrow Y^{[M]}$ (resp. $Z^{[M]}$). 
The limit density functions $f^{[M]}$ (resp. $g^{[M]}$) of $Y^{[M]}$ 
(resp. $Z^{[M]}$) is given by; \\
\textbf{\textrm{Case (A)}}: 
\begin{equation}\label{Mfixdens}
 f^{[M]}(x)=\sum_{j=0}^{[M/2]} 
	\binom{M}{j}\frac{\rho(x_j)}{|1-2j/M|}\mathcal{P}(x_j)^{M-j}\mathcal{Q}(x_j)^j
        + cI_{(M=even)}\delta_0(dx),
\end{equation}    
with $x_j=x/|1-2j/M|$, 
\begin{equation}
 \mathcal{P}(x) = \left(1+\sqrt{1-2x^2}\right)/2, \;\;
 \mathcal{Q}(x) = \left(1-\sqrt{1-2x^2}\right)/2, 
\end{equation} 
where $c$ is determined by 
\[ \int_{-\infty}^{\infty}f^{[M]}(x)dx=1. \]
\textbf{\textrm{Case (B)}}:  
\begin{equation}\label{Mfixdens B}
 g^{[M]}(x)=\left(\frac{1}{2}\right)^{M}\sum_{j=0}^{[M/2]} 
	\binom{M}{j}\frac{\rho(x_j)}{|1-2j/M|}+ c'I_{(M=even)}\delta_0(dx),
\end{equation} 
where $x_j=x/|1-2j/M|$ and $c'$ is determined by 
\[ \int_{-\infty}^{\infty}g^{[M]}(x)dx=1. \]    
\item \label{th-dfix}If we fix $d\geq 2$ and take $M \to \infty$, then 
$X_t^{[M]}/t \Rightarrow Y_d$ with the initial qubit of Case (A), 
while $X_t^{[M]}/\sqrt{t}\Rightarrow Z_d$ with the initial qubit of Case (B). 
Let $\mu_d(k)=\langle \widehat{\Psi}_d^{(\varphi_0)}, D_k\widehat{\Psi}_d^{(\varphi_0)}\rangle/d$, and 
$\nu_d(k)=\mathrm{Tr}[\widehat{H}^{-d}(k)\cdot D_k^2\widehat{H}^d(k)]/2d$ with $D_k=id/dk$. 
The limit distributions $Y_d$ and $Z_d$ are given by \\
\textbf{\textrm{Case (A)}}: 
\begin{equation} \label{dfixA} P(Y_d\leq x) = \int_{\{k\in [0,2\pi):\mu_d(k)\leq x\}}\frac{dk}{2\pi}, \end{equation}
\textbf{\textrm{Case (B)}}:
\begin{equation} \label{dfinxB} 
P(Z_d\leq x) = \int\int_{\{(k,u)\in [0,2\pi)\times \mathbb{R}: u(\nu_d(k))^{1/2}\leq x\}}
								\frac{e^{-u^2/2}dudk}{(2\pi)^{3/2}}.
\end{equation}
\end{enumerate}
\end{proposition}
\noindent
To prove Proposition 2, we will use the following lemma. 
Let the eigenvalues and eigenvectors of $\widehat{H}(k)$ be denoted by $\lambda_0(k)$, $\lambda_1(k)$ 
and $|v_0(k)\rangle$, $|v_1(k)\rangle$, respectively. 
\begin{lemma} \label{rem1}
 \begin{enumerate}
  \item \label{rem1-2}
  Let $h_j(k)=D_k(\lambda_j(k))/\lambda_j(k)$,\;($j \in \{0,1\}$). Then
  	\[ h_0(k)+h_1(k)=0. \] 
 \item \label{rem1-3}
  Define $p(k)=|\langle v_0(k),\varphi_0\rangle|^2$ and $q(k)\equiv 1-p(k)=|\langle v_1(k),\varphi_0\rangle|^2$. Then 
        \[ p(k)+q(k)=1. \]
 \item \label{konno}
 Let 
  $k(x)=\arccos(x/\sqrt{1-x^2})$. 
 Then 
 \[ \int_{0}^{2\pi}g(h_{j}(k))w(k)\frac{dk}{2\pi}=\int_{-\infty}^{\infty}g(x)w(k(x))\rho(x)dx.\;\; (j\in\{0,1\})\]
 \end{enumerate}
\end{lemma}
\noindent
From now on, we will prove Proposition \ref{prop2}. \\

\noindent\textbf{Proof of Proposition \ref{prop2}.} 
\begin{enumerate}
\item
Let $\lambda_j(k)=e^{i\theta_j(k)}$ $(j=0,\;1)$. We should remark that 
$(\theta_j(k+\xi/d)-\theta_j(k))d=\xi h_j(k)+O(d^{-1})$ and 
$|v_j(k+\xi/d) \rangle\langle v_j(k+\xi/d)|=|v_j(k) \rangle\langle v_j(k)|+O(d^{-1})$. 
By Lemma \ref{rem1} (\ref{rem1-2}) and (\ref{rem1-3}), we have for fixed $M$, 
\begin{align}
Q_d(k,\xi/t) 
&= e^{i\xi(h(k)+O(t^{-1}))/M}p(k)+e^{-i\xi (h(k)+O(t^{-1}))/M}q(k)+O(t^{-1}), \label{chara inf} \\
C_d(k,\xi/t) 
&= \frac{1}{2}\left(e^{i\xi(h(k)+O(t^{-1}))/M}+e^{-i\xi (h(k)+O(t^{-1}))/M}\right)+O(t^{-1}),\label{chara inf2}
\end{align}
where $h(k)\equiv h_0(k)$. 
Combining Lemma \ref{chara} with Eq. (\ref{chara inf}), the characteristic function with the 
initial qubit of Case (A) is described as 
\begin{equation}\label{Mfix}
\lim_{t \to \infty}E\left(e^{i\xi X_{t}^{[M]}/t}\right)
        =\sum_{j=0}^{M}\binom{M}{j}\int_{0}^{2\pi} e^{i\xi(1-2j/M) h(k)} p^{M-j}(k)q^j(k) \frac{dk}{2\pi}.    
\end{equation} 
From Lemma \ref{rem1} (\ref{konno}), 
we obtain the desired conclusion. We can also prove Case (B) by combining Lemma \ref{chara} with Eq. (\ref{chara inf2}). 
\item Noting that $D_k\widehat{H}(k)=\sigma_3\widehat{H}(k)$, 
we have $\mathrm{Tr}(\widehat{H}^{-d}(k)\cdot D_k\widehat{H}^{d}(k))=0$ for all $d$, where $\sigma_3$ is the Pauli matrix: 
\[ \sigma_3=\begin{bmatrix}1&0\\0&-1\end{bmatrix}. \]
Then we obtain for fixed $d$, 
\begin{align} 
Q_d(k,\xi/t^\theta) &= 1+i\frac{\xi}{t^\theta/d}\mu_d(k)+O(t^{-2\theta}), \label{dfix0}\\
C_d(k,\xi/t^\theta) &= 1-\frac{\xi^2/2}{t^{2\theta}/d}\nu_d(k)+O(t^{-3\theta}), \label{dfix 1}
\end{align}
with $\theta>0$. 
By applying Eq. (\ref{dfix 1}) to Lemma \ref{chara}, in Cases (A) and (B), we see 
\begin{align}
 \lim_{t\to \infty}E\left[e^{i\xi X_t^{[M]}/t}\right]
	&= \int_{0}^{2\pi} e^{i\xi \mu_d(k)}\frac{dk}{2\pi}, \label{charax}\\
 \lim_{t\to \infty}E\left[e^{i\xi X_t^{[M]}/\sqrt{t}}\right] 
	&= \int_{0}^{2\pi} e^{-\frac{\xi^2}{2}\nu_d(k)}\frac{dk}{2\pi}, \label{charax2}      
\end{align}
respectively. 
Thus Eq. (\ref{charax}) gives the desired conclusion in Case (A). 
For Case (B), noting that $e^{-\tilde{\xi}^2/2}$ is the characteristic function of $\mathrm{N}(0,1)$, 
Eq. (\ref{charax2}) can be rewritten as 
\[ \lim_{t\to \infty}E\left[e^{i\xi X_t^{[M]}/\sqrt{t}}\right]
	= \int_{0}^{2\pi}\int_{-\infty}^{\infty} e^{i\xi u(\nu_d(k))^{1/2}}\frac{dudk}{(2\pi)^{3/2}}. \]
\qquad\qquad\qquad\qquad\qquad\qquad\qquad\qquad\qquad\qquad\qquad\qquad\qquad\qquad\qquad\qquad\quad$\Box$
\end{enumerate}
\quad In the case of the initial qubit $|1\rangle^{\otimes M}$ for Proposition \ref{prop2} (1) and (2), 
we can obtain $E[(Y^{[M]})^2]=1-5/(4\sqrt{2})+1/(4M\sqrt{2})$ 
and $E[Z_d^2]=1/8$ ($d=2$), $7/72$ ($d=3$) in a similar way. 
The results were shown in Brun \textit{et al}. \cite{Brun}. \\
\quad In Proposition \ref{prop2} (2) ($d=2$ case), we have $\mu_2(k)=(\sin 2k)/2$, $\nu_2(k)=1$. 
Therefore we obtain the following corollary.
\begin{corollary}\label{cor}
As $t\to \infty$ with $t=2M$, in Case (A) (resp. (B)), $X^{[M]}/t\Rightarrow Y_2$ (resp. $Z_2$), where 
$Y_2$ has a scaled arcsine law distribution (resp. $\mathrm{N}(0,1)$). 
The limit density function of $Y_2$ is given by 
\[ 2s(2x)=\frac{2I_{(-1/2,1/2)}(x)}{\pi\sqrt{1-4x^2}}, \]
where $s(x)$ is defined by Eq. (\ref{arcsine}). 
\end{corollary}

\section{A crossover from classical behavior to quantum one }
As Brun \textit{et al.} claimed in \cite{Brun0}, the behavior of 
$M$-CQW with the initial qubit of Case (A) remains ``quantum'' 
in contrast with \cite{Brun2,Zhang} in the long time limit. 
However by considering the initial qubit of Case (B) and time steps $t\neq 0\;(\mathrm{mod}\;M)$, 
we can see a ``classical'' property. 
From Propositions \ref{prop1} and \ref{prop2} (\ref{th-dfix}) with the initial qubit of Case (A), 
the behavior of a particle becomes classical as $t\to \infty$ with $t\leq M$, 
while the behavior grows quantum with $t\geq 2M$ 
in the view point scaling order. 
So we will consider long time limit with $M<t<2M$. 
By Proposition \ref{prop2} (\ref{th-dfix}), the scaling order of the weak convergence 
is given by $t$ and $\sqrt{t}$ in Cases (A) and (B), respectively. 
To find a behavior corresponding to a kind of quantum to classical transition, 
we will introduce another initial qubit, that is, a mixture of Cases (A) and (B). 
From Proposition \ref{prop2} (\ref{th-Mfix}) and (\ref{th-dfix}) with the initial 
qubit of Case (B), the scaling order grows $t$ for fixed $M$, 
while it grows $\sqrt{t}$ for fixed $d$ in the long time limit. So we analyze the limit theorem for 
$d,M\to\infty$ simultaneously with the initial qubit of Case (B). 
Therefore, we consider the following three assumptions. Let $0<\beta<1$. \\
\\
\textbf{Assumption (a)}.  $t=M+M^\beta$ with the initial qubit of Case (A). \\
\textbf{Assumption (b)}.  $t=2M$ with a mixture of initial qubits of Cases (A) and (B), i.e., 
$\Phi_0=\varphi_0^{\otimes M^{\beta}}\otimes \phi_1\otimes\cdots\otimes\phi_{M-M^{\beta}}$, 
where $\{\phi_j\}_{j=1}^{M-M^\beta}$ is an i.i.d. sequence with $\phi_j=|1\rangle$, $|\!\!-\!\!1\rangle$ 
with probability $1/2$, respectively. \\
\textbf{Assumption (c)}.  $M \sim t^{1-\beta}$, $d \sim t^{\beta}$ with the initial qubit of Case (B), 
where $f(x)\sim g(x)$ means $\lim_{x \to \infty}f(x)/g(x)=1$. \\
\\
We obtain a phase diagram in relation to the limit distribution in each case;
Define 
$D=\{(\beta,\theta)\in [0,1]^2: \theta>\mathrm{max}\{1/2,\;\beta \}\}$, 
$D'=\{(\beta,\theta)\in [0,1]^2: \theta>(1+\beta)/2\}$. 
Let the scaled arcsine law distribution with the density function $2^\beta s(2^\beta x)$ 
be denoted by $F^{(\beta)}$. 
\begin{theorem}\label{main}
Let $\mathcal{X}$, $\mathcal{Z}$, and $\mathcal{W}^{(\beta)}$ be independent random variables 
with $\mathcal{X}\sim \mathrm{N}(0,1)$, $\mathcal{Z}\sim K$, and 
$\mathcal{W}^{(\beta)}\sim F^{(\beta)}$, 
where the distribution $K$ has the density function $\rho(x)$ defined by Eq. (\ref{Konnodens}). 
As $t \to \infty$, 
\begin{enumerate}
\item Under Assumption (a), 
\begin{equation*} 
X_t^{[M]}/t^\theta \Rightarrow 
	\begin{cases} \delta_0(x) & \text{if $(\beta,\theta)\in D$}, \\
                      \mathcal{X} & \text{if $\theta=1/2$ and $0\leq \beta<1/2$}, \\
                      \mathcal{X}+\mathcal{W}^{(0)} & \text{if $\theta=\beta=1/2$}, \\
                      \mathcal{W}^{(0)} & \text{if $1/2<\theta=\beta<1$}, \\
                      \mathcal{W}^{(1)} & \text{if $\theta=\beta=1$}.  	
\end{cases}
\end{equation*}
\item Under Assumption (b),
\begin{equation*}
X_t^{[M]}/t^{\theta} \Rightarrow
	\begin{cases} 
        	      \delta_0(x) & \text{if\;$(\beta,\theta)\in D$}, \\
                      \mathcal{X} & \text{if\;$\theta=1/2$ and $0\leq \beta<1/2$}, \\
                      \mathcal{X}+\mathcal{W}^{(1/2)} & \text{if\;$\theta=\beta=1/2$}, \\
                      \mathcal{W}^{(\beta)} & \text{if\;$1/2<\theta=\beta\leq 1$}. 
	\end{cases}
\end{equation*}        
\item Under Assumption (c),
\begin{equation*}
X_t^{[M]}/t^{\theta} \Rightarrow
	\begin{cases} 
        	      \delta_0(x) & \text{if\;$(\beta,\theta)\in D'$}, \\
                      \mathcal{X} & \text{if\;$\theta=1/2$ and $\beta=0$}, \\
                      \mathcal{X}\mathcal{Z} & \text{if\;$0<\theta<1/2$ and $\theta=(1+\beta)/2$}, \\
                      \mathcal{Z} & \text{if\;$\theta=1$ and $\beta=1$}.
	\end{cases}
\end{equation*}
\end{enumerate}
\end{theorem}
\noindent 
Under Assumptions (a) and (b), 
as long as $\beta<1/2$, the behavior is classical, while if $\beta>1/2$, then the standard deviation of a particle 
is proportional to $t^{\beta}$ and its limit distribution is described as a scaled arcsine law distribution. 
The cross point of the two regions, that is, $\beta=\theta=1/2$, the limit distribution is obtained 
by the convolution of $\mathrm{N}(0,1)$ and an arcsine law distribution. 
Under Assumption (c), 
the classical (resp. quantum) behavior appears at the end point $\beta=0$ (resp. $\beta=1$). 
If $0<\beta<1$, the scaling order 
of the weak convergence is 
$t^{(1+\beta)/2}$ and its limit distribution is expressed as product of two independent 
variables, $\mathcal{X}$ and $\mathcal{Z}$.
\begin{proof}
\begin{enumerate}
\item
Lemma \ref{chara} implies that the characteristic function for $X_t/t^{\theta}$ under Assumption (a) is 
\begin{equation}\label{chara1}
 E\left(e^{i\xi X_{t}^{[M]}/t^{\theta}}\right)
	= \int_{0}^{2\pi}Q_2(k,\xi/t^\theta)^{M^\beta}Q_1(k,\xi/t^\theta)^{M-M^\beta}\frac{dk}{2\pi}. 
\end{equation}
Remark that $\mu_1(k)=0$, $Q_1(k,\xi/t^\theta)=1-\xi^2/(2t^{2\theta})+O(t^{-3\theta})$. 
From the estimation and Eq. (\ref{dfix0}), as $t \to \infty$, 
\begin{equation}\label{q1c2}
Q_1(k,\xi/t^\theta)^{t} \to 
			      \begin{cases} 1 & \text{if\;$\theta >1/2, $} \\
					  e^{-\xi^2/2} & \text{if\;$\theta=1/2. $} 
                              \end{cases}
\end{equation}
\begin{equation}\label{q2}
Q_2(k,\xi/t^\theta)^{t^\beta} \to \begin{cases} 1 & \text{if\;$\theta <\beta, $} \\
					  e^{i\xi\mu_2(k)} & \text{if\;$\theta=\beta. $} 
                              \end{cases}
\end{equation}
When $\theta=\beta=1/2$, we see that by applying Eqs. (\ref{q1c2}) and (\ref{q2}) to Eq. ({\ref{chara1}}) 
\begin{align*} 
\lim_{t\to\infty}E\left(e^{i\xi X_{t}^{[M]}/t^{\theta}}\right)
	&= \int_{0}^{2\pi}\int_{-\infty}^{\infty}e^{i\xi(u+\sin 2k)}\frac{e^{-u^2/2}}{(2\pi)^{3/2}}dkdu, \\
        &= \int_{-\infty}^{\infty}\int_{-\infty}^{\infty}e^{i\xi(u+v)}\frac{e^{-u^2/2}}{\sqrt{2\pi}}s(v)dvdu,
\end{align*}
since $e^{-\xi^2/2}$ is the characteristic function of $\mathrm{N}(0,1)$ and $\mu_2(k)=\sin 2k$. 
For other cases except $\theta=\beta=1$, 
the desired conclusion can be derived similarly. 
Moreover Corollary \ref{cor} (Case A) gives the result of $\theta=\beta=1$. 
Therefore, we complete the proof of part (1). \\              
\item
The characteristic function under Assumption (b) can be expressed as 
\begin{equation} \label{opapi1}
E\left(e^{i\xi X_{t}^{[M]}/t^{\theta}}\right)
 	= \int_{0}^{2\pi}Q_2(k,\xi/t^\theta)^{M^\beta}C_2(k,\xi/t^\theta)^{M-M^\beta}\frac{dk}{2\pi}. 
\end{equation}
By Eq. (\ref{dfix 1}), we get 
\begin{equation} \label{opapi2}
C_2(k,\xi/t^{\theta})\to \begin{cases} 1 & \text{if\;$\theta >1/2, $} \\
					  e^{-\xi^2/2} & \text{if\;$\theta=1/2. $} 
                              \end{cases}
\end{equation}
From Eqs. (\ref{opapi1}) and (\ref{opapi2}), we have the desired conclusion 
in a similar fashion of part (1). 
\item 
Under Assumption (c), Lemma \ref{chara} (Case (B)) implies 
\begin{equation}\label{chara3}
 E\left(e^{i\xi X_{t}^{[M]}/t^{\theta}}\right)
	= \int_{0}^{2\pi}C_{d}(k,\xi/t^\theta)^{M}\frac{dk}{2\pi}. 
\end{equation}
Eq. (\ref{chara inf2}) yields 
\[
C_d(k,\xi/t^{\theta})=1-\frac{\xi^2}{2}h^2(k)t^{2(\beta-\theta)}+\mathrm{o}(t^{2(\beta-\theta)}),
\]
in the condition of $\beta-\theta<0$. 
So we have as $t\to\infty$, 
\begin{equation}\label{von}
\left\{C_{t^{\beta}}(k,\xi/t^\theta)\right\}^{t^{1-\beta}} \to 
				\begin{cases} 1 & \text{if\;$\theta>(1+\beta)/2, $} \\
					  e^{-\xi^2h^2(k)/2} & \text{if\;$\theta=(1+\beta)/2. $} 
                                \end{cases}
\end{equation}
Note that 
\begin{align}\label{stan}
e^{-\xi^2h^2(k)/2} &= \int_{-\infty}^{\infty} e^{i\xi h(k)u} \frac{e^{-u^2/2}}{\sqrt{2\pi}}du.
\end{align}
Applying Lemma \ref{rem1} (\ref{konno}) and Eqs. (\ref{von}) (\ref{stan}) to Eq. (\ref{chara3}), we have 
for $\theta=(1+\beta)/2$, 
\begin{align*} \lim_{t\to \infty}E\left(e^{i\xi X_{t}^{[M]}/t^{\theta}} \right)
        &=\int_{-\infty}^{\infty}\int_{-\infty}^{\infty} e^{i\xi uv} \rho(u)\frac{e^{-v^2/2}}{\sqrt{2\pi}}dudv.
\end{align*} 
Proposition \ref{prop1} and Proposition \ref{prop2} (\ref{th-Mfix}) for $M=1$ give 
results of $(\beta,\theta)=(0,1/2)$ and 
$(\beta,\theta)=(1,1)$. Therefore we complete the proof of part (3). 
\end{enumerate}
\end{proof}

\noindent Under assumption $t=dM$, Eqs. (\ref{chara inf2}) and (\ref{dfix 1}) imply that for $0<\beta<1$, 
\begin{align*}
\lim_{t\to\infty}\left\{C_{t^{\beta}}\left(k,\xi/t^{\frac{1+\beta}{2}} \right)\right\}^{t^{1-\beta}}
 &=\lim_{M\to\infty}\lim_{d\to\infty}\left\{C_d\left(k,\frac{\xi}{\sqrt{M}d}\right)\right\}^M \\
 &=\lim_{d\to\infty}\lim_{M\to\infty}\left\{C_d\left(k,\frac{\xi}{\sqrt{M}d}\right)\right\}^M \\
 &=e^{-\xi^2h^2(k)/2}. 
\end{align*} 
Similarly, from Eqs. (\ref{chara inf}) and (\ref{dfix0}), we see that for $0<\beta<1$, 
\begin{align*} 
\lim_{t\to\infty}\left\{Q_{t^{\beta}}\left(k,\xi/t \right)\right\}^{t^{1-\beta}}
 &=\lim_{M\to\infty}\lim_{d\to\infty}\left\{Q_d\left(k,\frac{\xi}{Md}\right)\right\}^M \\
 &=\lim_{d\to\infty}\lim_{M\to\infty}\left\{Q_d\left(k,\frac{\xi}{Md}\right)\right\}^M \\
 &=e^{i\xi h(k)(p(k)-q(k))}. 
\end{align*}
Therefore, applying the above equations to Lemma \ref{chara}, we get the following result. 
\begin{proposition}
Let $t=dM$. We impose the same assumtion in Proposition \ref{prop2}.
\begin{enumerate}
\item $\sqrt{M}Z^{[M]}\Rightarrow \mathcal{X}\mathcal{Z}$ ($M\to \infty$), and 
$Z_d\Rightarrow \mathcal{X}\mathcal{Z}$ ($d\to \infty$). 
\item Assume that $d \sim t^{\beta}$ and $M \sim t^{1-\beta}$ with initial qubit of Case (A). Then we have 
$X_t^{[M]}/t\Rightarrow Y$ as $t\to \infty$, for $0<\beta<1$, 
where $Y$ has the density function 
\[ f(x)=\frac{3I_{(-1/\sqrt{8},1/\sqrt{8})}(x)}{\pi(1+x^2)\sqrt{1-8x^2}} .\]
Furthermore, 
$Y^{[M]}\Rightarrow Y$ ($M\to \infty$), and 
$Y_d \Rightarrow Y$ ($d\to \infty$). 
\end{enumerate}
\end{proposition}
\noindent In the case of the initial qubit $|1\rangle^{\otimes M}$ for part (2), 
we can obtain the limit distribution $Y$ similarly, 
where $Y$ has 
the density function 
\[ f(x)=\frac{I_{(0,1/2)}(x)}{\pi (1-x)\sqrt{(1-2x)x}}. \]
This gives $E[Y^2]=1-5/(4\sqrt{2})$ which was shown by Brun \textit{et al.} \cite{Brun}.





\section*{Acknowledgments}
We would like to thank Makoto Katori for useful discussions and comments.








\end{document}